# Numerical Evaluation of Fragility Curves for Earthquake-Liquefaction-Induced Settlements of an Embankment

C. Khalil, A.M. ASCE[1], I. Rapti, Ph.D.[1] and F. Lopez-Caballero, Ph.D.[1]

[1]Laboratoire MSS-Mat CNRS UMR 8579, CentraleSupélec Paris-Saclay University, Grande Voie des Vignes, Chatenay-Mâlabry, 92290 France, e-mail : christina.khalil, ioanna.rapti, fernando.lopez-caballero@centralesupelec.fr

**ABSTRACT**

The major cause of earthquake damage to an embankment is the liquefaction of the soil foundation that induces ground level deformations. It is well known that the liquefaction appears when the soil loses its shear strength due to the excess of pore water pressure. This phenomenon leads to several disastrous damages of the soil foundation. The aim of this paper is to assess numerically the effect of the liquefaction-induced settlement of the soil foundation on an embankment due to 76 real earthquakes extracted from the PEER database. For this purpose, a 2D finite element model of a dam founded on a layered soil/rock profile was considered. An elastoplastic multi-mechanism model was used to represent the soil behaviour. The crest settlement of the embankment was selected as the quantifiable damage variable of the study. Fragility functions were drawn to give the probability exceedance of some proposed damage levels as function of a seismic severity parameter. In addition, the anisotropy was tested by the change in the soil permeability and a comparison with the isotropy was held. According to the results, the crest settlement increases with the peak ground acceleration and the fragility functions showed that above 0.2g, the probability to have moderate damage in the anisotropic case reaches unity whereas it is lesser in the isotropic case. The embankment will not show serious damage for this same value of acceleration in the two cases.
<u>Keywords</u>: *liquefaction, damage levels, anisotropy, fragility functions, crest settlement*

**INTRODUCTION**

Earthquakes are the most natural phenomenon that cause damage to the soil and to the structures, in addition to other losses such as human and economic losses. *Liquefaction* phenomenon is considered as one of the most devastating and complex behaviors that affect the soil due to shakings. It is defined as the loss of the soil of its shear strength due to the excess of pore water pressure. The most affected structures by liquefaction foundation are the earth dams (Siyahi et al. 2008; Wu 2014). To best design a dam, its stability and performance should be taken into account and they depend on many factors.

There are several ways of failure of a dam: a disruption by a fault movement, a slope failure, a piping failure through cracks or a crest settlement. The crest settlement is the parameter that easily quantifies the damage failure of an earth dam. According to Swaisgood (2003), the damages could be divided into four levels based on the crest settlement of the embankment and the peak ground acceleration of the input signal. In addition to the soil disruption, taking into account the anisotropic case will lead to almost real estimation of the soil behaviour. Hence, the consideration of the soil variability and uncertainties by the change in the permeability will encounter larger settlements.

The following paper aims to assess numerically the effect of soil liquefaction-induced failure to a dam due to real earthquakes. It is based on the Performance Based Earthquake Engineering methodology (PBEE) developed by the federally funded earthquake engineering research center (Pacific Earthquake Engineering Research PEER). A deterministic study to quantify a failure way of a dam (crest settlement) and a probabilistic study to find the probability of exceedance of a certain level of performance, took place. Fragility functions were drawn for this purpose. Hence, two stages of the PEER methodology were satisfied. In order to account for the natural hazards, the input ground motions were used and chosen to be real motions to be consistent with the seismic parameter, magnitude, site to source distance, design and the duration of the earthquake (Wu 2014). The finite element calculations were performed using the GEFDyn software. A comparison between the isotropic and anisotropic cases was analyzed.

**MODEL DESCRIPTION**

GEOMETRY AND FE MODEL
The geometry of the model, as shown in Figure 1, consists of an embankment of 9m high composed of dry dense sand. The soil foundation is composed by a liquefiable loose sand of 4m at the top of a saturated dense sand of 6m. The bedrock at the bottom of the dense sand is 5m and has the shear wave velocity $V_s$ = 1000 m/s. The water table is situated at 1m below the base of the dam and the dam was kept dry. The dam's inclination is a slope of 1:3 (vertical: horizontal).

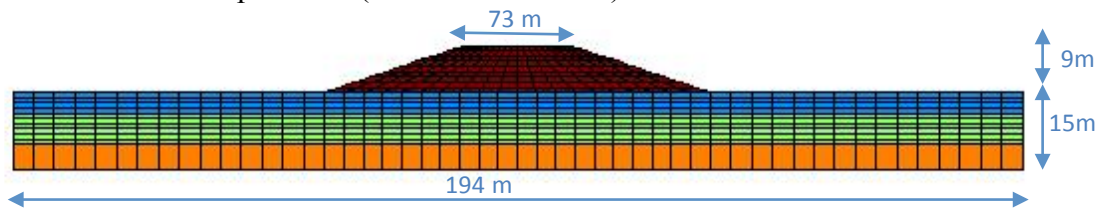

**Figure 1: Illustration of the geometric model**

A 2D coupled finite element modelling with GEFDyn Code (Aubry et al. 1986) is carried out using a dynamic approach derived from the u-$p_w$ version of the Biot's generalized consolidation theory (Zienkiewicz and Taylor, 1991) was adopted for the soil. The FE model is composed of quadrilateral isoparametric elements (3.7m x 1m) with eight nodes for both solid displacements and fluid pressures. Each element has 9 integration points. So the total number of nodes is 2331 nodes. The time step of the calculation is set to be $10^{-3}$s. The soil permeability for the loose sand is taken as $k=10^{-4}$m/s and is supposed to be isotropic. The FE analysis is performed in three consecutive steps: i) a computation of the initial in-situ stress state due to gravity loads; ii) a sequential level-by-level construction of the embankment and iii) a seismic loading analysis in the time domain.

For the boundary conditions of the static phase, the horizontal displacement is blocked at the lateral surface of the meshing whereas the vertical displacement is allowed. For the base of the meshing, only the vertical displacement is not allowed. Concerning the dynamic phase, only vertically incident shear waves are introduced into the domain and as the response of an infinite semi-space is modelled, equivalent boundaries have been imposed on the nodes of lateral boundaries. For the half-space bedrock's boundary condition, paraxial elements simulating "deformable unbounded elastic bedrock" have been used (Modaressi and Benzenati 1994).

SOIL BEHAVIOR MODEL

The ECP elastoplastic multi-mechanism model (Aubry et al. 1982; Hujeux 1985) is used to represent the soil behavior of three types of sand under cyclic loading. The non-linearity of this model is represented by four-coupled elementary plastic mechanism: three plane-strain deviatoric plastic strain mechanism in three orthogonal planes (k-planes) and an isotropic plane to take into account normal forces. The detailed study of this model is beyond the scope of this work but for further examination, a check on the following research of Lopez Caballero et al. (2007) and Lopez Caballero et al. (2010) would be helpful. The soil model parameters were determined with the procedure defined by Lopez-Caballero et al. (2007).

INPUT GROUND MOTION

The selection of input motions for geotechnical earthquake engineering problems is important as it is strongly related to the nonlinear dynamic analyses. For the scope of this study, 76 real earthquakes were selected. They were chosen to be real earthquakes in order to represent the consistency of the seismic parameters and characteristics. Most of them were extracted from the PEER database and are classified into 5 groups:

- NF = Near Fault (20 signals); these motions have a strong velocity pulse so they attribute important losses and damages.
- PL = Pulse Like (20 signals); Baker (2007) provides a pulse indicator to identify the pulse like motions taking into consideration the characteristics of the motion.

- NPL = Non Pulse Like (20 signals); these signals have a small velocity pulse and a longer duration pulse.
- LA = Low Amplitude (10 signals) extracted from kik-Net network in Japan; this nomination only designates that the signals have low amplitudes and does not refer to any classification.
- Other = (6 signals); they were chosen arbitrary to give a variety for the study.

## DETERMINISTIC ANALYSIS

The PEER methodology deals with four stages: the hazard analysis in which an intensity measure (IM) parameter is identified, the structural analysis in which the response to the earthquake is represented by the engineering demand parameter (EDP), the damage analysis in which the probability of failure is quantified and the final stage is the loss analysis which requires the estimation of the decision based on the cost and maintenance of the project. This work would be dealing with two stages of this methodology: the structural and the damage analysis. The structural analysis requires a determistic approach to calculate the used parameters of the study. Concerning the damage analysis, the crest settlement induced by the liquefaction apparition is defined as the EDP which must be linked with the IM parameter. In this section, the variation of the excess pore water pressure ($\Delta p_w$) and the vertical displacement $u_z$ are calculated during the co-seismic time. For the sake of brevity, only two signals were chosen to conduct the deterministic analysis: signal_1 and signal_2. They were chosen arbitrary in a way that the first has smaller peak acceleration than the second. The accelerograms of these selected motions are shown in Figure 2.

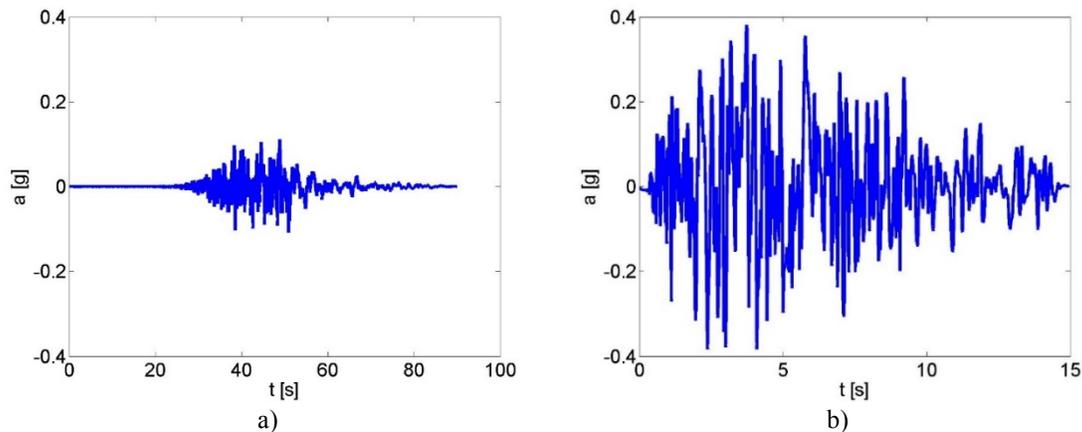

**Figure 2: Accelerograms of the selected motions: a) signal_1 and b) signal_2**

DISTRIBUTION OF EXCESS PORE WATER PRESSURE

The distribution of the excess pore water pressure is studied in order to identify the liquefaction phenomena during the signal duration. It was first examined with respect to the co-seismic time for the two selected earthquakes. The distribution of the excess pore water pressure is studied with respect to the horizontal distance far from the dam. For that purpose, three nodes at the same level were chosen (refer to Figure 3 to see the nodes locations). Their depth is 2m down in the soil and their coordinates are 26, 37 and 57 to the left of the dam center as shown in Figure 3. Note that the origin of the axis is the base center of the dam. Then three relative nodes were chosen at the depth of 7m. For the first motion, the results shown in Figure 4, indicate that with respect to the horizontal distance far from the dam, the excess pore water pressure increases progressively during the shaking and it shows high values under the dam and small values far away. Then during the mainshock or around it, the excess pore water pressure has a peak value and starts to decrease afterwards to approach zero. It is clear that the liquefaction happens because the soil starts to lose its shear strength.

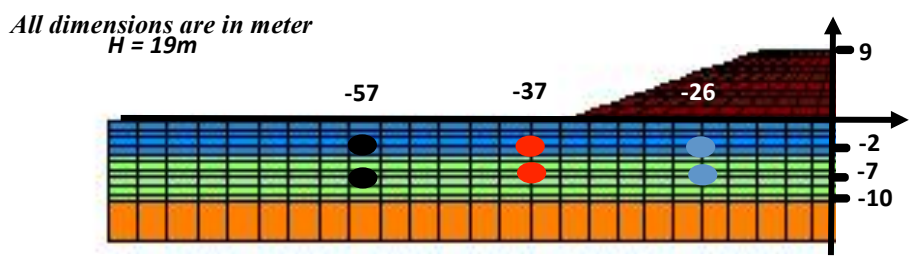

**Figure 3: Location of the selected nodes**

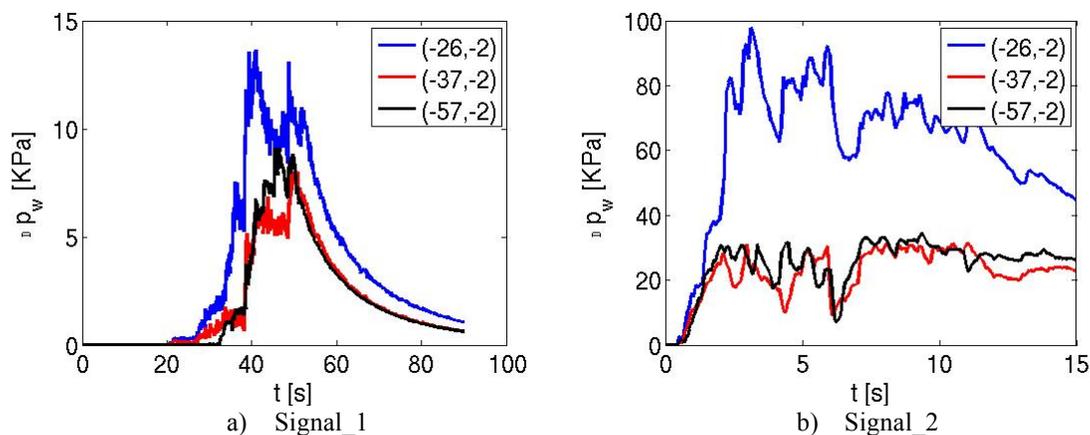

a) Signal_1  b) Signal_2
**Figure 4: Distribution of excess pore water pressure during the selected signals**

For signal_1, the generation of excess pore water pressure reaches its maximum of 13 kPa at the mainshock for the closest node to the dam and after the mainshock for the other tested nodes. $\Delta p_w$ decreases as far as we get from the dam. A similar behavior is found for signal_2, and due to high acceleration value, the $\Delta p_w$ at 2m deep and below the embankment is close to 100kPa. It can be partially concluded that the distribution of the excess pore water pressure is higher at the surface where the loose sand is placed.

CREST SETTLEMENT

For dams under seismic activities, the modes of failure usually studied are the crest settlement or the internal erosion and piping failure caused by cracks in the dam (Wu 2014). In this study, the crest settlement is chosen to be the mode of failure because it is a quantifiable measurement. The obtained co-seismic settlement for the two input signals is shown in Figure 5 for the crest as well as at the free field. For the crest, the settlement increases rapidly in order to reach a value at the mainshock after which it continues to be constant for signal_1 but continues to increase slowly for signal_2. Hence, the crest settlement is higher in signal_2 than in signal_1. Notice that in this particular case, the signal duration did not affect the result because signal_2 has smaller duration but generates more displacement.

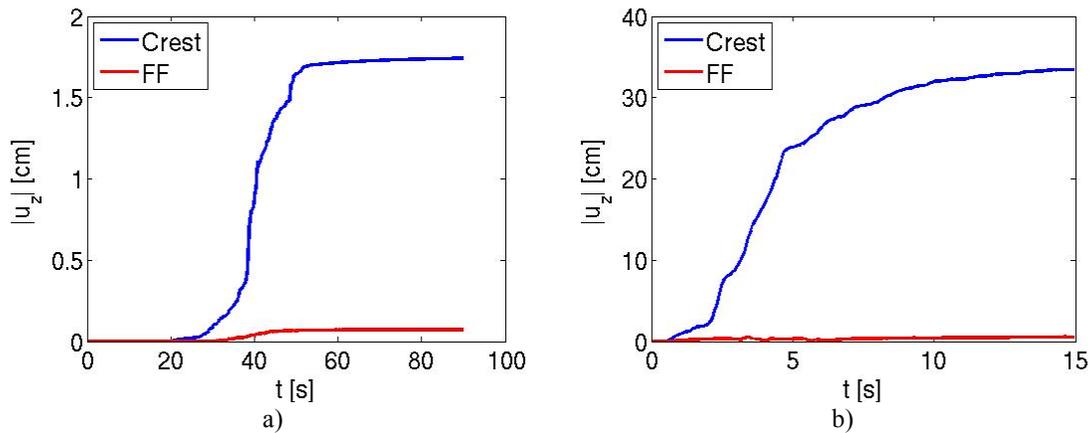

**Figure 5: Vertical displacement of a) signal_1 and b) signal_2**

Swaisgood (2003) analyzes a historical database on the performance of dams during earthquakes and found that the crest settlement is directly related to some input ground motion characteristics (i.e. the peak ground acceleration and magnitude). Following Swaisgood's proposition, in this work the obtained percentage crest settlement ($\frac{\delta u}{H}$, where u is the crest settlement, H is the height of the dam and the foundation which is 19m as seen in figure 3) is compared to the peak ground acceleration at the outcropping bedrock ($a_{max\ out}$). To take into account all the tested signals, the crest settlement was calculated accordingly and was drawn as function of $a_{max\ out}$ (Figure 6). It is interesting to note that, according to Figure 6, the calculated crest settlement increases when the acceleration at the outcrop increases.

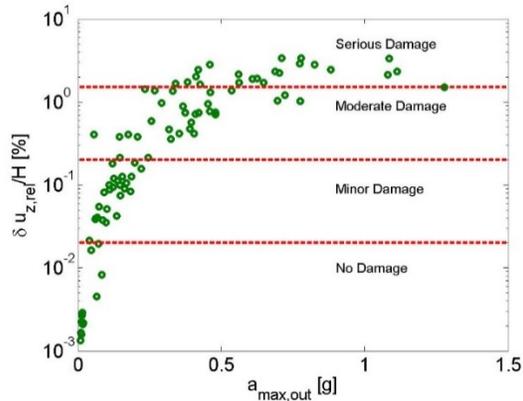 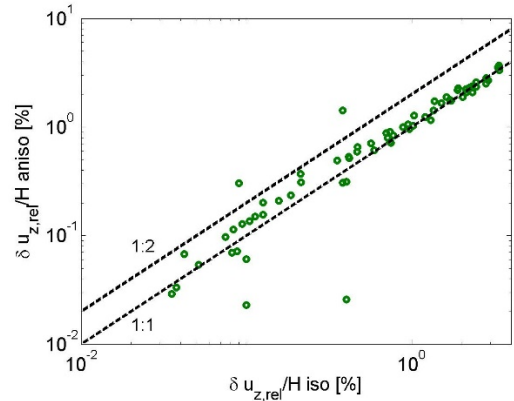

Figure 6: Obtained induced crest settlement as a function of peak ground acceleration.  Figure 7: Comparison of the crest settlement between the isotropic and anisotropic case

ANISOTROPIC CASE

According to Witt (1983), three facts can lead to permeability anisotropy: macro-stratification, micro-stratification and orientation and flatness of the particles. In addition, soil properties vary in all directions based on several parameters such as the depositions of particles, the weathering or the physical environment (Sanchez Lizarraga et al. 2014). Also a particle size distribution or a particle shape examination would better explain the reason because the permeability is related to the pore size as well as the particle elongation (Masad & Munhunthan, 2000). To account for this variability, a change in a soil parameter was taken into consideration: the permeability. So as to highlight the effect of anisotropy on the model response, the horizontal permeability was kept as $k_y=10^{-4}$m/s whereas the vertical permeability was changed to $k_z=10^{-6}$m/s. Since the crest settlement is the considered engineering demand parameter, it is calculated for the anisotropic case for all the tested input motions. A comparison between the obtained values in isotropic and anisotropic case is shown in Figure 7. It can be seen from Figure 7 that for higher values of induced settlement the effect of the anisotropy could be neglected. However, it is noted that for lower values of both settlement and input acceleration the change in permeability affects the pore water pressure dissipation and soil behaviour, which is traduced with an increase in settlements for the anisotropic case.

**DAMAGE ANALYSIS**

According to Porter (2003) among others, various analytical approaches to assess the level performance of a certain structures are developed. As an example, the Load Resistance Factor design approach assures the performance based on the failure probability of the individual structure whereas the Performance Based Earthquake Engineering (PBEE) evaluates the performance based on the risk of collapse. In addition, to account for the global geotechnical uncertainties presented in the soil that are affected by the seismic activities, the context of the PBEE would best serve for a probabilistic analysis. Such analysis is conducted to determine the probability of various levels of damage analysis as well as the examination that a certain level of damage will exceed a certain limit investigated by fragility curves (Popescu 2005; Saez et al 2010).

Hence, in the context of the PBEE, the damage analysis, which is the third stage of this methodology, is a procedure to quantify the structural damage. It consists of setting fragility functions in order to find the probability of the design to exceed a certain level of performance. As mentioned in the previous section, for the scope of this research, the crest settlement induced by the liquefaction apparition will be considered as the engineering demand parameter (EDP). Hence, $a_{max\ out}$ could be used as the intensity measure (IM) to predict the evolution of this EDP in the studied embankment. A comparison between the isotropic and anisotropic case is held also. From Swaisgood's (2003) work, four damage levels related to the percentage of crest settlement were proposed and they are shown as red dashed lines in Figure 6.

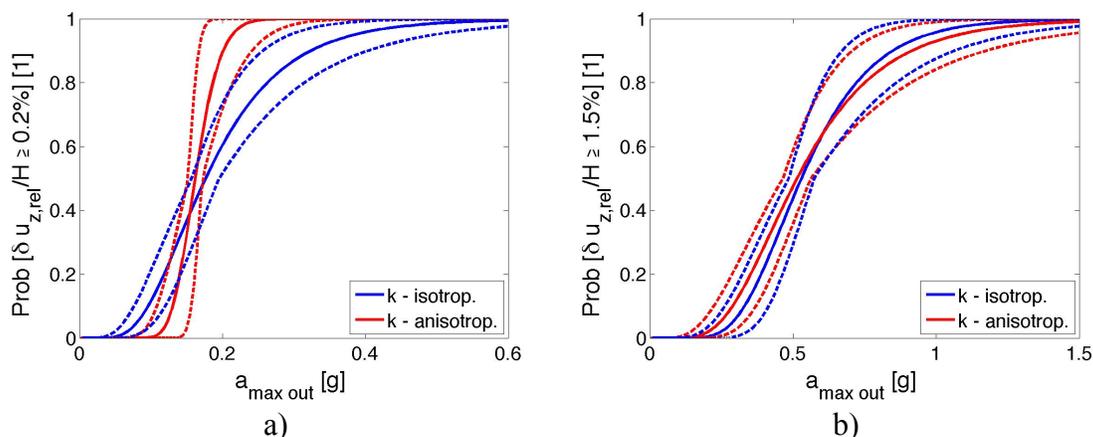

**Figure 8: Fragility function based on the percentage crest settlement for a) moderate damage level and b) serious damage level**

Using the 76 motions, the maximum likelihood method was used to compute numerical values of the estimators of parameters defining the fragility curve under the lognormal assumption. The obtained fragility curves for the third and fourth state damages (minor to moderate and moderate to serious damages) are shown in Figure 8. These curves are drawn as solid lines whereas the statistical confidence of the derived fragility curves are drawn as dashed lines.

It can be seen from Figure 8.a) that the probability of failure for the moderate damage level in the isotropic case is reached before the anisotropic case. For acceleration of 0.2g for example, there is 65% chance that the embankment will generate moderate damage in the isotropic case whereas it will fail in the anisotropic case. For this acceleration also, the embankment will not show serious damage regardless of the case (Figure 8.b). Notice that in Figure 8.b), the two cases show slightly close values. These results confirm that the effect of the anisotropy in the permeability is important for the estimation of the minor to moderate damages and that it could be neglected for the case of serious damages.

**CONCLUSION**

A numerical assessment of the soil liquefaction induced settlement for an embankment dam due to real earthquakes was presented in this paper. An elastoplastic multi-mechanism soil behaviour model was used with the help of a 2D finite element code (GEFDyn). The Performance Based Earthquake Engineering methodology was investigated from which two stages were held: the structural analysis and the damage analysis. The engineering demand parameter for this study is the induced crest settlement since it is a quantifiable parameter that is important when dealing with the type of failures of a dam.

First, the analysis was conducted deterministically to calculate the distribution of the pore water pressure and the induced settlement of the model. The study took into account the natural hazards so the 76 real input motions were implemented in the analysis. Notice that the earthquakes were extracted mainly from the PEER database. The anisotropy of the model was taken into consideration by the change in the horizontal permeability and a comparison between the results was done as well.

For the deterministic analysis, the results show that the signal with higher acceleration and hence more severity, leads to a complete deterioration of the embankment. For the damage analysis, the crest settlement is proportional with the acceleration at the outcrop. On the other hand, when dealing with anisotropy, the results show that for some signals there is no influence of the permeability on the model whereas for others the soil behaves differently in each case. The crest settlement was divided into four categories based on it the probabilistic analysis was conducted. Fragility functions were drawn for that purpose.

Finally, we can say that the results obtained are compatible with previous studies conducted with almost similar cases. Further research can be done to ameliorate the results as to account for the influence of permeability on the soil and even the variability of the soil properties or the variation of the water table.